\documentclass[letter,twocolumn]{aa}

\usepackage[T1]{fontenc}
\usepackage[utf8]{inputenc}

\usepackage{graphicx}
\usepackage{natbib}
\usepackage{ulem}
\usepackage{textcomp}
\usepackage{gensymb}
\usepackage{longtable}
\usepackage{threeparttable}
\usepackage{multicol}
\usepackage{multirow}
\usepackage{textgreek}
\usepackage{float}
\usepackage{todonotes}
\usepackage[Symbol]{upgreek}
\usepackage{amsmath}
\usepackage{etoolbox}
\usepackage{txfonts}
\usepackage{url}
\usepackage{xcolor}

\usepackage[most]{tcolorbox}
\usepackage{hyperref}
\usepackage{xcolor}
\hypersetup{
  colorlinks   = true, 
  urlcolor     = blue, 
  linkcolor    = blue, 
  citecolor   = blue 
}
\usepackage[all]{hypcap} 

\begin{document} 
	\title{A perfect power-law spectrum even at highest frequencies: The Toothbrush relic}
	\titlerunning{A power-law spectrum from 58 MHz to 18.6 GHz}
	\authorrunning{Rajpurohit et al.}

\author{K. Rajpurohit\inst{1,2,3}, F. Vazza\inst{1,2,4}, M. Hoeft\inst{3}, F. Loi\inst{5}, R. Beck\inst{6}, V. Vacca\inst{5}, M. Kierdorf\inst{6}, R. J. van Weeren\inst{7} , D. Wittor\inst{3},  F. Govoni\inst{5}, M. Murgia\inst{5}, C. J. Riseley\inst{1,2}, N. Locatelli\inst{1}, A. Drabent\inst{3}, and  E. Bonnassieux\inst{1}}

\institute{Dipartimento di Fisica e Astronomia, Universit\'at di Bologna, via P. Gobetti 93/2, 40129, Bologna, Italy\\
 {\email{kamlesh.rajpurohit@unibo.it}}
\and
INAF-Istituto di Radio Astronomia, Via Gobetti 101, 40129 Bologna, Italy
\and
Th\"uringer Landessternwarte (TLS), Sternwarte 5, 07778 Tautenburg, Germany
\and
Hamburger Sternwarte, Universit\"at Hamburg, Gojenbergsweg 112, 21029, Hamburg, Germany
\and
INAF-Osservatorio Astronomico di Cagliari, Via della Scienza 5, 09047 Selargius (CA), Italy
\and
Max-Planck-Institut f\"ur Radioastronomie, Auf dem H\"ugel 69, 53121 Bonn, Germany
\and
Leiden Observatory, Leiden University, P.O. Box 9513, 2300 RA Leiden, The Netherlands
}
   
     \abstract
  {Radio relics trace shock fronts generated in the intracluster medium (ICM) during cluster mergers. The particle acceleration mechanism at the shock fronts is not yet completely understood. We observed the Toothbrush relic with the Effelsberg and Sardinia Radio Telescope at 14.25\,GHz and 18.6\,GHz, respectively. Unlike previously claimed, the integrated spectrum of the relic closely follows a power law over almost three orders of magnitude in frequency, with a spectral index of $\alpha_{\rm 58\,MHz}^{\rm 18.6\,GHz}=-1.16\pm0.03$. Our finding is consistent with a power-law injection spectrum, as predicted by diffusive shock acceleration theory. The result suggests that there is only little magnetic field strength evolution downstream to the shock. From the lack of spectral steepening, we find that either the Sunyaev-Zeldovich decrement produced by the pressure jump is less extended than $\sim$ 600\,kpc along the line of sight or, conversely, that the relic is located far behind in the cluster. For the first time, we detect linearly polarized emission from the ``brush'' at 18.6\,GHz. Compared to 8.3\,GHz, the degree of polarization across the brush increases at 18.6\,GHz, suggesting a strong Faraday depolarization towards lower frequencies.  The observed depolarization is consistent with an intervening magnetized screen that arise from the dense ICM containing turbulent magnetic fields. The depolarization, corresponding to a standard deviation of the Rotation Measures as high as $\sigma_{\rm RM}= 212\pm23\rm \,rad\,m^{-2}$, suggests that the brush is located in or behind the ICM. Our findings indicate that the Toothbrush can be consistently explained by the standard scenario for relic formation.}


   \maketitle


\section{Introduction}
\label{sec:intro}
Radio relics are large, diffuse sources that are associated with powerful shock fronts originating in the intracluster medium (ICM) during clusters merger \citep[for a review, see e.g.][]{Feretti2012,vanWeeren2019}. One striking observational feature of radio relics is their high degree of polarization. The magnetic field vectors are often found to be well aligned with the shock surface \citep{vanWeeren2010,Bonafede2012,Owen2014,DeGasperin2014,Kierdorf2016}.

Despite progress in understanding radio relics, the actual acceleration mechanism at the shock fronts is not fully understood. It is generally believed that diffusive shock acceleration \citep[DSA;][]{Drury1983} generates the observed cosmic ray electrons (CRe). However, it is currently debated if the acceleration starts from the thermal pool \citep[standard scenario;][]{Ensslin1998,Hoeft2007} or from a population of mildly relativistic electrons \citep[re-acceleration scenario;][]{Kang2011,Kang2016a}

\begin{figure*}
    \centering
    \includegraphics[width=1.0\textwidth]{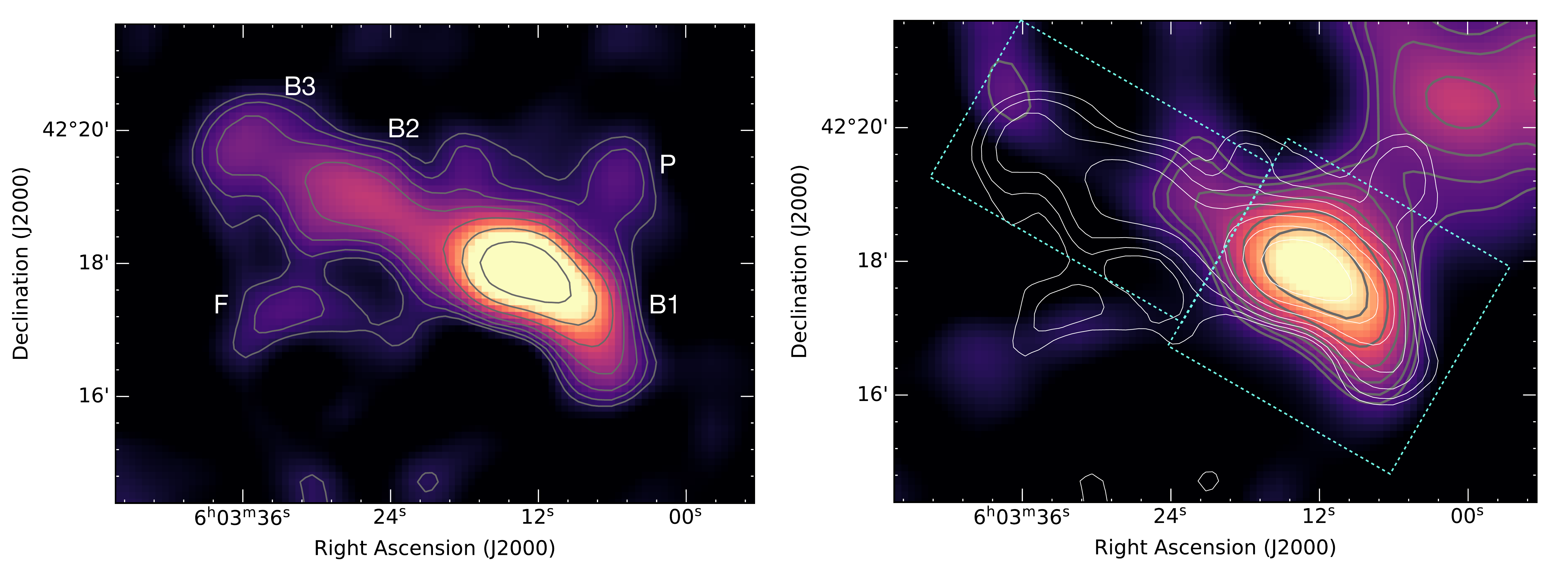}
        \vspace{-0.5cm}
 \caption{Total power emission from the Toothbrush relic at 70\arcsec resolution. $\textit{Left}$: Effelsberg 14.25\,GHz image. The largest linear size of the relic is $\sim1.8\,\rm Mpc$, similar to those reported below 10\,GHz. Contour levels are drawn at $\sqrt{[1,2,4,8,\dots]}\,\times\,3\,\sigma_{{\rm{ rms}}}$, where $\rm \sigma_{rms, 14.25\,GHz}=0.2\,m\,Jy\,beam^{-1}$ and  $\rm \sigma_{rms,18.6\,GHz}=0.4\,m\,Jy\,beam^{-1}$. $\textit{Right}$: SRT 18.6\,GHz image overlaid with the SRT (gray) and Effelsberg (white) contours. Cyan boxes define the area used for measuring the integrated spectrum of the relic and its sub-regions. The emission at the top right corner in the SRT image is due to blending of discrete sources.}
      \label{fig1}
  \end{figure*}

The standard scenario has successfully reproduced many of the observed properties of relics, however, three major difficulties remain: (i) the spectra of some relics are reported to show a spectral break above 10\,GHz \citep{Stroe2016}, which is incompatible with the power-law spectrum predicted by DSA theory, (ii) a power-law energy distribution from the thermal pool CRe energies relevant for the synchrotron emission may require an unphysical acceleration efficiency \citep{vanWeeren2016a,2019arXiv190700966B}, and (iii) the Mach numbers derived from X-ray observations are often significantly lower than derived from the overall radio spectrum \citep{Akamatsu2012,2019arXiv190700966B}. 

According to the re-acceleration scenario the shock fronts re-accelerate electrons from a pre-existing fossil population. There are a few examples, which seem to show a connection between the relic and active galactic nuclei. \citep{Bonafede2014,vanWeeren2017a,Gennaro2018,Stuardi2019}. If relics originate according to the re-acceleration scenario, weak shocks may become radio bright, solving issue (ii) and (iii). A break in the radio spectrum is expected at high frequency, when the shock passes through a finite size cloud of fossil electron population \citep{Kang2016a}. If the fossil population is homogeneously distributed, the re-acceleration scenario also predicts a power-law spectrum.

The merging galaxy cluster 1RXS\,J0603.3+4213, located at redshift $z=0.225$, is one the most intriguing clusters hosting a spectacular toothbrush-shaped relic \citep{vanWeeren2012a,vanWeeren2016a,Rajpurohit2018,Rajpurohit2019,Gasperin2020}. It consists of three distinct components, namely the brush (B1) and two parts forming the handle (B2+B3). The relic shows an unusual linear morphology and is quite asymmetric with respect to the merger axis. The handle extends into very low density ICM. 
 
\citet{Stroe2016} reported evidence for a spectral steepening above 2.5\,GHz in the integrated radio spectrum of the relic. This claim was mainly based on the 16\,GHz and 30\,GHz radio interferometric observations. It has been suggested that the steepening in the integrated radio spectrum can be reproduced with the re-acceleration scenario \citep{Kang2016b}. \citet{Basu2016} studied the impact of the Sunyaev-Zeldovich (SZ) effect on the observed synchrotron flux density. They suggested that SZ contamination leads to a high frequency steepening for relics, albeit not at the level claimed by \citet{Stroe2016}. Recently, we studied the integrated spectrum of the relic between 120\,MHz to 8\,GHz and excluded any steepening up to 8\,GHz \citep{Rajpurohit2019}. However, the spectral behavior of the relic remained uncertain between 10-20\,GHz. The Toothbrush is known to be highly polarized \citep{vanWeeren2012a}. Effelsberg observations revealed a high fractional polarization at 8.3\,GHz and a strong depolarization and Rotation Measure (RM) gradient from the brush to the handle \citep{Kierdorf2016}. 

The main aim of this paper is to answer the question if the overall spectrum of the Toothbrush steepens in the frequency range between 10-20\,GHz. If the spectrum steepens at high frequency, this would have a tremendous impact on the radio relic formation scenario, since it would clearly be in conflict with the standard scenario for relic formation, which predicts a power law towards high frequencies. A steepening would be difficult to explain within the standard scenario and would favor the re-acceleration scenario. We adopt a flat $\Lambda$CDM cosmology with $H_{\rm{ 0}}=70$ km s$^{-1}$\,Mpc$^{-1}$, $\Omega_{\rm{ m}}=0.3$, and $\Omega_{\Lambda}=0.7$. At the cluster's redshift, $1\arcsec$ corresponds to a physical scale of 3.64\,kpc.

\section{Observations}
The radio observation at 14.25\,GHz were performed with the Effelsberg 100-m telescope with the new  Ku-band
receiver in dual polarization mode. The total on-source observation time was 20\,hours  with 2500\,MHz bandwidth. We obtained 31 coverages of a field of $11\times7\,\rm arcmin^{2}$ and processed the data with the NOD3 tool \citep{Muller2017}. The data reduction involves Radio Frequency Interference removal and baselevel corrections, like Basket-Weaving of two maps with scanning in orthogonal directions (RA/DEC).

\begin{figure*}
    \centering
    \includegraphics[width=0.49\textwidth]{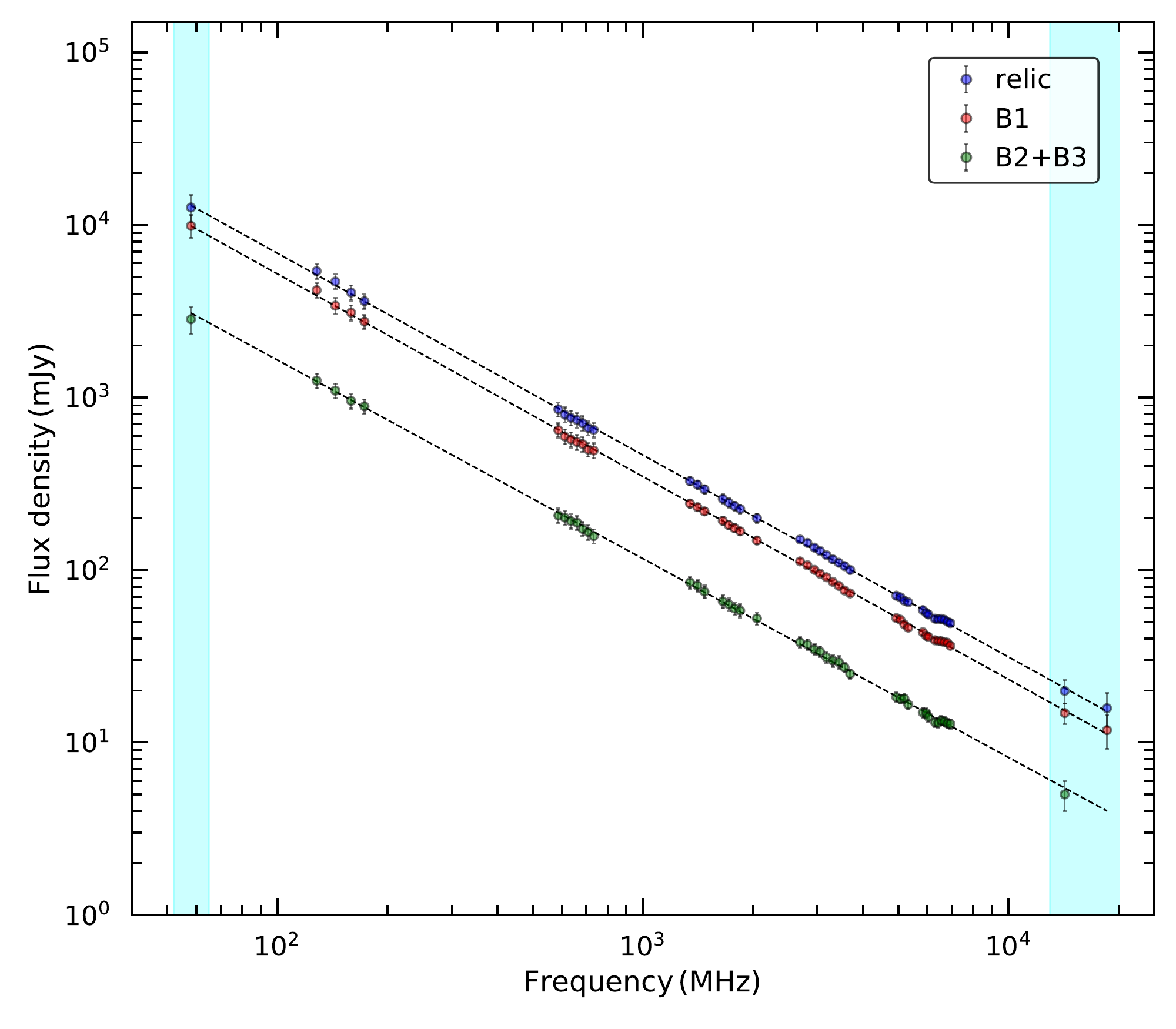}
     \includegraphics[width=0.49\textwidth]{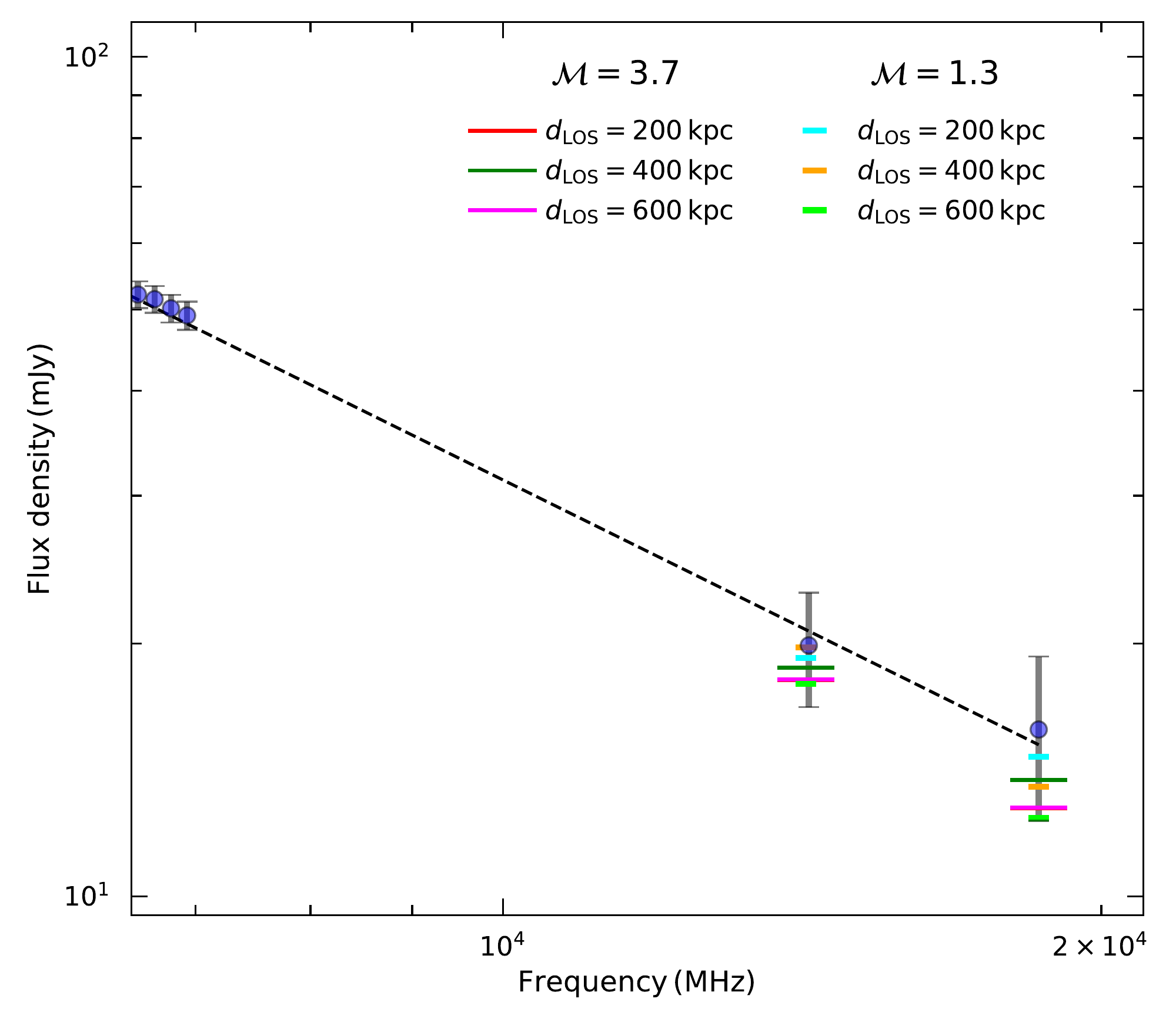}
    \vspace{-0.3cm}
 \caption{\textit{Left}: Integrated spectrum of the Toothbrush relic between 58\,MHz and 18.6\,GHz. Dashed lines show the fitted power laws. The spectrum follows a close power law with a slope of $\alpha=-1.16\pm0.03$. The new flux density points are highlighted by the cyan regions, other values are adopted from \citet{Rajpurohit2019}. \textit{Right}: The possible impact of SZ decrement (shown with horizontal color lines) as a function of line of sight depth ($d_{\rm LOS}$) on the radio spectra of the relic emission. Blue circles show the observed flux densities. In order to produce an SZ decrement compatible within the error-bars of the new 14.25 and 18.6\,GHz observations, the depth of the shock pressure jump along the line of sight is required to be $d_{\rm LOS} \leq 600 \rm ~kpc$.}
      \label{fig2}
\end{figure*}

The Sardinia Radio Telescope (SRT) observations were performed in a full polarization mode with the 7-feed K-Band receiver centered at 18.6\,GHz with a bandwidth of 1200\,MHz. The observations were carried out between January and February 2020, for a total of 24\,hours. The data were reduced using the proprietary software package Single-dish Spectral-polarimetry Software \citep[SCUBE;][]{Murgia2016}.  

The uncertainty in the flux density measurements were estimated as: 
\begin{equation}
  \Delta{S_{\nu}}
  =
  \sqrt{(f\cdot S_{\nu})^2+ N_{\rm beam}(\sigma_{\rm rms})^{2}},
\end{equation}
where $f$ is the absolute flux density calibration uncertainty, $S_\nu$ is the flux density, $\sigma_{{\rm{ rms}}}$ is the rms noise and $N_{{\rm{beams}}}$ is the number of beams. We assume an absolute flux density uncertainty of 10\,\% for both SRT and Effelsberg.


\section{Results and discussion}
In Figure\,\ref{fig1}, we show the Effelsberg and the SRT total intensity images at 70\arcsec resolution. The relic is clearly detected at both frequencies. The largest linear size of the relic is $\sim1.8\,\rm Mpc$, similar to the one reported below 10\,GHz. We measure the flux density of $19.9\pm3.1$\,mJy and $15.8\pm3.5$\,mJy at 14.25\,GHz and 18.6\,GHz, respectively. These values are significantly larger than those reported by \citet{Stroe2016}, namely $S_{16\,\rm GHz}=10.7\pm0.8\,\rm mJy$. We speculate that the discrepancy between our measurements and the one taken by the Arcminute Microkelvin Imager interferometer is due to the ``resolved-out'' effects. Interferometric observations underestimate the flux density of extended emission when the size of emission region gets close to Largest Angular Scale detectable with the interferometer.

\subsection{Integrated spectrum}
To obtain the integrated spectrum of the relic, we combine our new flux density measurements with those presented in \cite{Rajpurohit2019}. In addition, we include the flux density measurements from the LOFAR LBA observations at 58\,MHz \citep{Gasperin2020}.  We measure flux densities for the entire relic as well as for the regions B1 and B2+B3; see Table\,\ref{tabl1}.

The resulting integrated spectra are shown in the left panel of Figure\,\ref{fig2}. We find that the relic follows a close power law over almost three orders of magnitude in frequency. 
The integrated spectral index of the relic between 58\,MHz and 18.6\,GHz is $-1.16\pm0.03$. The spectral index value is consistent with our previous estimates \citep{Rajpurohit2018,Rajpurohit2019}. Recently, the power-law spectrum results are also found for the relic in CIZA\,J2242.8+5301 \citep{Loi2020}. The power-law spectrum is consistent with the standard scenario for the relic formation. Conversely in the framework of the re-acceleration scenario, the absence of an upper frequency spectral break would imply that a finite size cloud of fossil electron population is very large and distributed homogeneously. As a result, we do not see its effect in the relic overall spectrum.

According to the DSA theory in the test-particle regime and adopting a constant shock strength and CRe cooling in a homogeneous medium, the ``integrated'' spectrum is related to the Mach number according to
\begin{equation}
    \mathcal{M}
    =
    \sqrt{\frac{\alpha_{\rm int} -1}{\alpha_{\rm int} +1}}
    .
    \label{eq::alpha_int_mach}
\end{equation}
The radiative lifetime of electrons observed at 58\,MHz is about 120\,Myr when adopting a magnetic field strengths of 1\,$\mu$G. If the shock propagates with 1000\,km\,s$^{-1}$ this corresponds to length of about 160\,kpc. The slope of the spectrum down to 58\,MHz can only interpreted as a Mach number according to Eq.~\ref{eq::alpha_int_mach} if the physical conditions at the shock do not change significantly on a scales of 160\,kpc. This condition is likely fulfilled for the Toothbrush since it is located at a projected distance to the cluster center of about 1.1\,Mpc. The index above therefore corresponds to a Mach number of $\mathcal{M}=3.7\pm0.3$.

Despite the fact that the brush is about 4 times brighter than the handle, the entire relic and its sub-regions follow a power-law behavior and show similar spectral slopes; see Table\,\ref{tabl1}. At face value, this implies that the shock strength remains the same over $\sim1.8\,\rm Mpc$ scale. As argued in \cite{Rajpurohit2019}, the shock surface indeed shows a distribution of Mach numbers, thus a single Mach number derived above can only roughly characterize the shock. Most importantly, the tail of the Mach number distribution towards high values determine the radio spectral index \citep{2019arXiv190911329W,Rajpurohit2019}.

Our finding is  basically consistent with the standard scenario for the formation of radio relics if the radio spectral index corresponds to the Mach number of the shock. If the shock would have a strength as low as estimated from the X-ray surface brightness, no radio emission which could be detected with current telescopes is expected \citep{vanWeeren2012a,2019arXiv190700966B}. If the shock has instead a strength as estimated from the radio spectral index, the standard DSA-based scenario is in accord with the observations, if a strong magnetic field  and efficient electron acceleration is adopted \citep[see, e.g., Fig. 9 in][]{2019arXiv190700966B}. We note, however, that even in this situation a few percent of the kinetic energy dissipated at the shock front needs to be transferred by DSA to the supra-thermal accelerated at the shock front.

\subsection{Constraints on the downstream magnetic field evolution}

It is conceivable, that the magnetic field strength downstream of the shock increases, e.g., due to a turbulent dynamo process driven by the curvature of the shock front, or decreases, e.g., by expansion of the shock compressed material. Depending on frequency, the observed radio emission probes very different volumes. At the highest frequency, 50\,\% of the emission are emitted from a volume with an extent of about 5\,kpc downstream to the shock front. In contrast, the emission at 58\,MHz is extended to about 85\,kpc. If the strength of the magnetic field would change significantly on these lengths scales, this would affect the integrated spectrum of the relic. 

A non-linear change of the field strength would either significantly boost the emission at short or at large distances, in both cases this would result in a curved spectrum, see e.g., \citet{2016MNRAS.462.2014D}. Since the integrated spectrum almost perfectly follows a power law, only a marginal non-linear increase or decrease of the magnetic field strength seems to be possible on scales probed by the relic. 

However, if the field strength changes linearly with distance, the power-law integrated spectrum is preserved but the relation Equation~\ref{eq::alpha_int_mach} does not hold anymore. An increasing field strength would steepen the integrated spectrum while a decreasing one would flatten it. If the field strength doubles one a scale of 85\,kpc, the spectrum would steepen by about $-0.2$ (the actual value depends on many parameters as, e.g., the field strength itself). If the relic is formed according to the standard scenario, such a steep magnetic field gradient is clearly disfavored  by the observations. A decreasing downstream field strength might be consistent with our observations, however, it would significantly aggravate the efficiency problem.

%
\setlength{\tabcolsep}{3pt}
\begin{table}[!htbp]
\caption{Integrated flux densities}
\centering
\begin{threeparttable} 
\begin{tabular}{ c c c c  c  c}
 \hline  \hline  
Region &$S_{58\,\rm MHz}$ &$S_{14.25\,\rm GHz}$ &$S_{18.6\,\rm GHz} $ & $\alpha_{58\,\rm MHz}^{18.6\,\rm GHz}$ \\
  &Jy &mJy & mJy &  \\
relic&$12.6\pm2.3$&$19.9\pm3.1$&$15.8\pm3.5$&$ -1.16\pm0.03$ \\ 
B1 &$9.8\pm1.5$&$15.1\pm2.0$ & $11.9\pm2.6$&$-1.17\pm0.03$\\
B2+B3&$2.8\pm0.5$& $4.8\pm0.8$&$-$&$-1.15\pm0.04$\\
\hline 
\end{tabular}
\begin{tablenotes}[flushleft]
\footnotesize
\item{\textbf{Notes.}} The flux densities 14.2 and 18.6\,GHz are measured from 70\arcsec resolution images. Flux density at 58 MHz are measured from LOFAR LBA image. The relic flux exclude the contribution from sources F and P.
\end{tablenotes}
\end{threeparttable} 
\label{tabl1} 
\end{table}


\subsection{SZ decrement between 10-20 GHz}
The SZ effect contributes a negative signal against the
cosmic microwave background for $\nu\leq 220\rm\,GHz$. In the case of relics, \citet{Basu2016} showed that the SZ effect from the shock downstream also scales proportional to the Mach number squared, producing a contamination within exactly the same spatial scales responsible for the relic emission.

At 15\,GHz, the SZ effect is expected to reduce the observed synchrotron flux density by $\sim 10-50\%$, and must be taken into account when attempting a physical interpretation in case of any deviation from the power-law spectra. Conversely, since the SZ decrement {\it must} be expected if the shock leading to observed relic involve thermal gas, a lack of spectral steepening can be used to further constrain the shock parameters.

The SZ decrement at a given observation frequency depends on the line-of-sight projection of the pressure jump, $d_{\rm LOS}$, and therefore on the (unknown) shock geometry at the location of the relic. For a simple plane-parallel geometry and ignoring curvature, the total SZ decrement can be obtained by integrating $y$ over the visible relic area: $L \times \cal W$, where $L$ is the shock length and $\cal W$ is the width, leading to an angular size of the relic  $\Omega_{\mathrm{relic}} \approx L{\cal W}/D_A^2$ steradians (with $D_A$ the angular diameter distance). 
Following \citet{Basu2016}, we calculate the maximum {\it total} allowed SZ flux decrement from the region sampled by our new high frequency radio observation of the Toothbrush:

\begin{equation}
\begin{split}
|\Delta S_{\nu,\rm relic}^{\mathrm{SZ}}| \leq {0.26} & ~\mu\mathrm{Jy}~ \left(\dfrac{D_A}{700~ \mathrm{Mpc}}\right)^{-2} 
	\left(\dfrac{L}{1~ \mathrm{Mpc}}\right) \left(\dfrac{d_{\rm LOS}}{1~ \mathrm{Mpc}}\right) \left(\dfrac{\cal W}{100~ \mathrm{kpc}}\right) \\
	& \times \left(\dfrac{n_{\mathrm{u}} T_{\mathrm{u}}}{10^{-4}~ \mathrm{keV~ cm}^{-3}}\right)~ 
	\left(\dfrac{M}{3}\right)^2 \left(\dfrac{\nu}{1.4~ \mathrm{GHz}}\right)^2 .
\end{split}	
\label{eq:Snusz}
\end{equation}

We use $D_A=751 \rm ~Mpc$, $L=1.86 \rm ~ Mpc$,  $\cal W$=$422$ kpc, and two possible shock strengths, either $\mathcal{M}=3.7$ (as suggested by radio observations) or $\mathcal{M}=1.3$ (as suggested by X-ray analysis, see \cite{Ogrean2013,vanWeeren2016a}). $n_u$ and $T_u$ are the pre-shock density and temperature that can be derived by the two Mach numbers, respectively. For each Mach number, the temperature and density are derived from the standard Rankine-Hugoniot jump conditions based on the assumed post-shock values, $n_d=3\times10^{-3}\rm\,cm^{-3}$and $T_d=6 \rm ~keV$ \citep{vanWeeren2016a}. 

We thus produce estimates of $|\Delta S_{\rm \nu,relic}^{\mathrm{SZ}}| $ for different frequencies, by fixing the above model parameters and varying the unknown value of $d_{\rm LOS}$. Our results are given in the right panel of Figure\,\ref{fig2}. In order to produce an SZ decrement compatible within the error-bars of our 14.25 and 18.6\,GHz observations, the depth of the shock pressure jump along the line of sight is required to be $d_{\rm LOS} \leq 600 \rm ~kpc$ for a shock of strength $\mathcal{M}=3.7$. For a shock of strength $\mathcal{M}=1.3$, the SZ decrement at $d_{\rm LOS} = 600 \rm ~kpc$ already produces  a spectrum falling below the error bars of our observations. Hence, requiring an even smaller depth of the shock along the line of sight.  
We emphasize that the quoted values only refers to the contribution to the SZ decrement from the shock discontinuity along the line of sight, for the same range of spatial scales responsible for the radio emission. 

Furthermore, the assumption of a simple planar geometry and the absence of curvature along the line of sight is clearly an oversimplification, which may indeed explain the surprisingly low value of $d_{\rm LOS}$. Incidentally, such a small SZ decrement may also be explained if the shock responsible for the relic is at a more peripheral location in the cluster. In this case the density and temperature values suggested by X-ray observations originate from regions which are denser than the one responsible for the radio emission. In this case, Equation\,\ref{eq:Snusz} would significantly overestimate the pressure jump at the shock, and the requirement on $d_{\rm LOS}$ would be relaxed. 

\begin{figure}
    \centering
    \includegraphics[width=0.5\textwidth]{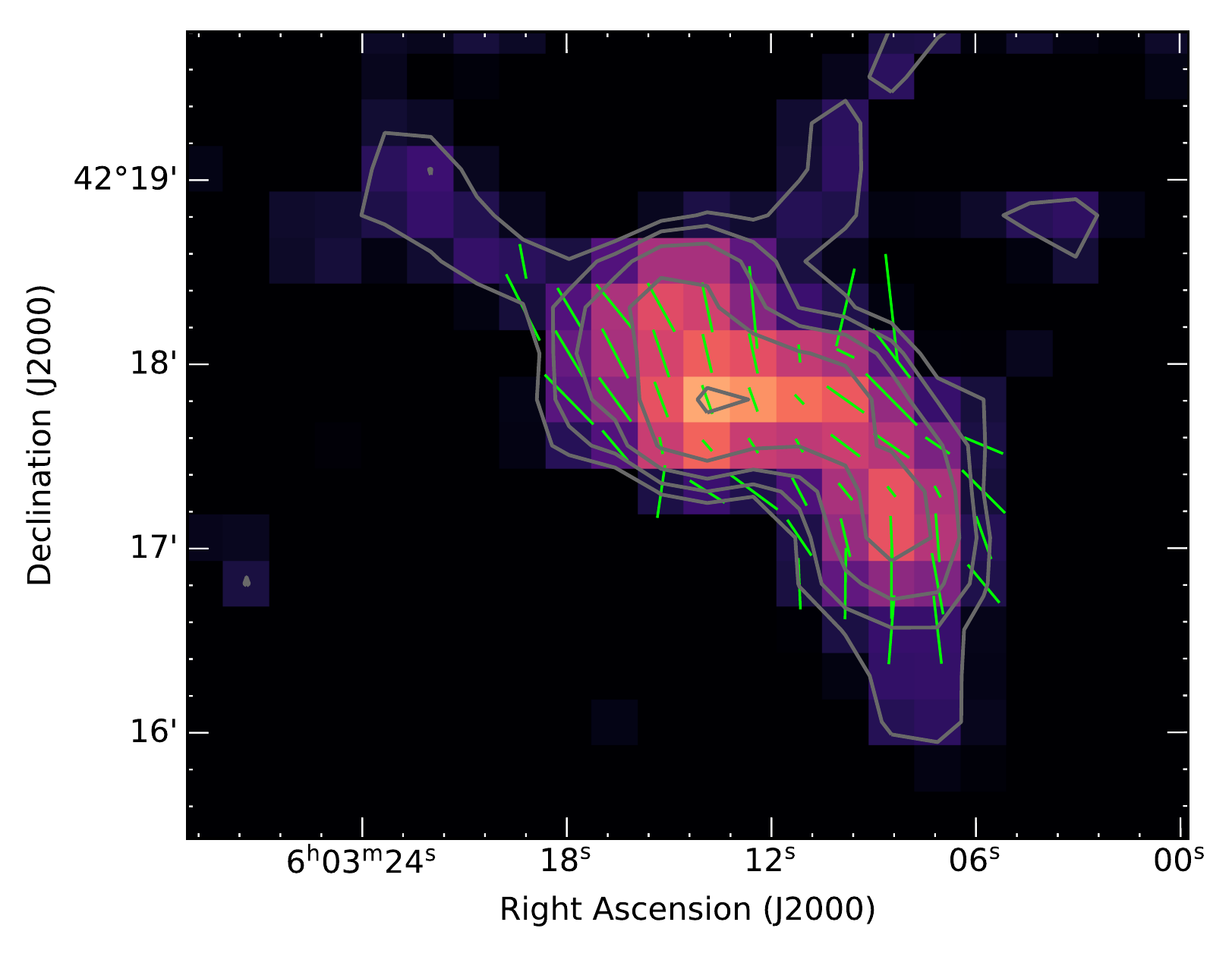}
    \vspace{-0.8cm}
 \caption{B-vectors distribution across the brush region at $51\arcsec$   resolution overlaid with the SRT total power contours at $3\sigma$. The length of the vectors depict the degree of polarization. The vectors are corrected for Faraday rotation effect. The mean polarization fraction at the brush is ($30\pm7)\%$.}
      \label{fig3}
  \end{figure}    
  

\subsection{Polarization at 18.6 GHz}
All of the information on the polarization properties of relics are mainly collected in the frequency range of 1-8.3\,GHz. Since the Faraday rotation is expected to be almost negligible at 18.6\,GHz, the intrinsic polarization of the relic could be directly mapped by our observations.

For the first time, we detect polarized emission from the relic at 18.6\,GHz. We detect polarized emission mainly from the brush region; see Figure\,\ref{fig3}. The degree of polarization varies along the brush and the magnetic field vectors are mainly aligned to the relic orientation. The fractional polarization reaches $\sim66\%$ in some areas, the average being $\sim30\pm7\%$. We note that these values could be affected by beam depolarization. 

Previous polarization measurements of the toothbrush relic have shown that the fractional polarization of B1 decreases rapidly towards lower frequencies. B1 is polarized at a level of about 15\% at 8.3\,GHz \citep{Kierdorf2016} and about 11\% at 4.9\,GHz. The polarization fraction drops below 1\% at frequency near 1.4\,GHz \citep{vanWeeren2012a}. The comparison between 8.3\,GHz and our measurement suggests significant depolarization even between 18.6 and 8.3\,GHz. Other than the Toothbrush relic, the polarization observations above 4.9\,GHz are available only for three relics, namely the Sausage relic, the relic in ZwCl\,0008+52, and Abell\,1612 \citep{Kierdorf2016, Loi2017}. For the above mentioned relics, the fractional polarization remains nearly constant at 4.9\,GHz and 8.3\,GHz.

The standard deviation of the RM, $\sigma_{\rm RM}$, is a useful parameter to characterize Faraday rotation and depolarization caused by an external Faraday screen. The depolarization induced by an external Faraday screen containing turbulent magnetic fields \citep{Burn1966,Sokoloff1998} can be described as
\begin{equation}
p(\lambda)= p_{0}\,e^{-2\sigma_{\rm RM}^{2} \lambda^{4}},
\label{sigmarm}
\end{equation}
where $p_{0}$ is the intrinsic polarization fraction. The maps between 4.9 and 18.6\,GHz show depolarization of $\rm DP_{4.9}^{18.6}=0.36\pm0.07$ for B1. This enables us to derive  $\sigma_{\rm RM}= 212\pm23\rm \,rad\,m^{-2}$. The observed $\sigma_{\rm RM}$ for the brush of Toothbrush is several times higher than for any other radio relic. 
This indicates that the brush region of the relic experiences strong Faraday rotation from the dense ICM. The strong depolarization suggests that the emission lies in or behind the ICM, which is very likely causing a low Mach number shock detected via X-ray observations \citep{Ogrean2013,vanWeeren2016a} .


\section{Conclusions}
We presented high frequency radio observations of the Toothbrush relic with the SRT and the Effelsberg telescope. We find that the relic follows a close power-law spectrum between 58\,MHz to 18.6\,GHz, with a slope of $\alpha=-1.16\pm0.03$. Our findings indicate that the Toothbrush can be consistently explained by the standard scenario for relic formation. The slope of the spectrum disfavors that the strength of the magnetic field significantly changes on scales probed by the radio emission, i.e., about 85\,kpc. 

We detected polarized emission at 18.6\,GHz. Compared to measurements at lower frequencies, the polarization fraction of the brush increases at 18.6\,GHz. The high value of $\sigma_{\rm RM}$ is consistent with $\sigma_{\rm RM}$ fluctuations of an ICM screen with tangled magnetic fields. This suggests that the brush is located in or behind the ICM. 

From the lack of steepening in the relic spectra, we find that either the SZ decrement at the shock along the line of sight is $\leq 600 \rm ~kpc$ thick, or the pressure jump associated with the relic is located far behind in the cluster. The latter explanation can also be reconciled with the trends of polarization fraction for the brush region.

\begin{acknowledgements}
KR and FV acknowledge financial support from the ERC Starting Grant ``MAGCOW'', no. 714196. FL acknowledge financial support from the Italian Minister for Research and Education (MIUR), project FARE, project code R16PR59747, project name FORNAX-B. RJvW acknowledges support from the VIDI research programme with project number 639.042.729, which is financed by the Netherlands Organisation for Scientific Research (NWO). CJR and EB acknowledges financial support from the ERC Starting Grant ``DRANOEL''number 714245. AD acknowledges support by the BMBF Verbundforschung under grant 05A17STA. We thank Sorina Reile for processing part of the Effelsberg data. Based on observations with the 100-m telescope of the MPIfR (Max-Planck-Institut f\"ur Radioastronomie) at Effelsberg. The Sardinia Radio Telescope \citep{Bolli2015,Prandoni2017} is funded by the Ministry of Education, University and Research (MIUR), Italian Space Agency (ASI), the Autonomous Region of Sardinia (RAS) and INAF itself and is operated as National Facility by the National Institute for Astrophysics (INAF). The development of the SARDARA back-end has been funded by the Autonomous Region of Sardinia (RAS) using resources from the Regional Law 7/2007 ``Promotion of the scientific research and technological innovation in Sardinia''.
\end{acknowledgements}

\bibliographystyle{aa}

\bibliography{ref.bib}

\end{document}